\documentclass[preprint,pra,showpacs,showkeys,amsmath,amssymb,floatfix]{revtex4-1}

\usepackage{graphicx}
\usepackage{dcolumn}
\usepackage{bm}
\usepackage{epsfig}
\usepackage{color}

\usepackage{pifont}
\usepackage{color}

\begin{document}

\title{Electronic and optical properties of metal-doped TiO$_2$ nanotubes: spintronic and photocatalytic applications}

\author{Mohamed M.\ Fadlallah$^{1,2}$}\
\author{Ulrich Eckern$^{3}$}\
\affiliation{
$^1$Physics Department, Faculty of Science, Benha University, 13518 Benha, Egypt\\
$^2$Center for Computational Energy Research, Department of Applied Physics, Eindhoven University of Technology, 
P.O. Box 513, 5600 MB Eindhoven, The Netherlands\\
$^3$Institute of Physics, University of Augsburg, 86135 Augsburg, Germany}


\begin{abstract}
Due to their characteristic geometry, TiO$_2$ nanotubes (TNTs), suitably doped by metal-substitution to enhance their
photocatalytic properties, have a high potential for applications such as clean fuel production. In this context, 
we present a detailed investigation of the magnetic, electronic, and optical properties of transition-metal 
doped TNTs, based on hybrid density functional theory. In particular, we focus on the $3d$, the $4d$, as well as 
selected $5d$ transition-metal doped TNTs. Thereby, we are able to explain the enhanced optical activity and 
photocatalytic sensitivity observed in various experiments. We find, for example, that Cr- and W-doped TNTs 
can be employed for applications like water splitting and carbon dioxide reduction, and for spintronic devices. 
The best candidate for water splitting is Fe-doped TNT, in agreement with experimental observations. In addition, 
our findings provide valuable hints for future experimental studies of the ferromagnetic/spintronic behavior of 
metal-doped titania nanotubes. 
\end{abstract}

\keywords{titania nanotubes; metal-doping; electronic and optical properties; photocatalytic and spintronic applications;
water splitting; clean fuel production}

\maketitle

\section{Introduction}
\label{sec:intro}

Titanium dioxide (TiO$_{2}$, titania) nanostructures have seen exciting applications in a number of areas, 
including batteries \cite{ba1}, sensors \cite{a2}, sunscreens \cite{c2}, photovoltaics \cite{b2}, solar cells \cite{s2}, 
biomedicals \cite{bio1}, catalyst supports \cite{cats2}, photocatalytic degradation of pollutants \cite{deg2}, 
carbon dioxide reduction \cite{co1}, hydrogen production \cite{hyd1}, and water splitting \cite{wat2}. In addition, doped 
TiO$_{2}$ nanostructures are promising materials for ferromagnetic and spintronic applications \cite{fer3,fer4}. 
Due to their low cost, natural abundance, high and long-term stability, and human and environmental safety,
titania is ubiquitous in daily life, e.g., in papers, inks, pigments \cite{paint,pigm}, 
toothpaste \cite{too}, cosmetics, medications, and food products \cite{foo}. 

In particular, a titania nanotube (TNT) is an effective nano-photocatalyst that directly splits water \cite{split1,split2},
and degrades environmental pollutants \cite{deg2} under sunlight. Furthermore, it is used in solar energy 
conversion \cite{In1} due to the good locations of its conduction band (CB) and valence band (VB) edges with respect 
to hydrogen formation and oxidation energy \cite{In2}. Moreover, the highly ordered nanotube geometry and large 
internal surface area are very useful as a unidirectional electric channel for the photogenerated electrons \cite{In5}. 
However, the bandgap of TNT (3.18--3.23 eV \cite{tnte1,tnte2}) restricts its applications in photocatalytic 
processes because of the limited absorption in the visible-light range. Therefore, engineering the bandgap of TNT by 
dopants to increase its photosensitivity to visible light is a major target in photocatalyst studies. On the 
other hand, Co- \cite{spintCo} and Ni-doped \cite{spintNi} TNTs can be used as dilute magnetic semiconductors. 

Most publications in the field have been concerned with mechanisms that decrease the bandgap of TNTs, and shift 
the absorption edges towards the visible-light range. Doping is a common method for tuning the bandgap of 
semiconductors. Experimentally, a large number of doped TNTs was prepared, e.g., Refs.~\cite{Lu2014,Y0,nanoimp,Zn00,Ru00};
and theoretical studies, based on density functional theory (DFT), include, e.g., Refs.~\cite{CNSFe,FADL,fadl2}. 

With respect to magnetic properties, ferromagnetic behavior has been experimentally observed for 
V- \cite{Vmag}, Ni- \cite{spintNi}, and Co-doped \cite {spintCo} TNTs. Theoretically, these dopants are found
to induce significant magnetic moments and long-range ferromagnetic coupling \cite{mag}.

In view of the relative simplicity of preparing metal-doped TNTs \cite{simple} (M-TNTs), further theoretical
studies are highly needed to elucidate systematically the photocatalytic and magnetic properties of such systems.
This paper is devoted to such a task, in particular, we present results of a systematic and accurate---based on
hybrid density funcional theory---investigation of the structural, electronic, and optical properties of 
3$d$, 4$d$, and selected 5$d$ metal-doped TNTs, in order to contribute to a better understanding of available
experimental results for photocatalytic and spintronic properties, as well as for providing hints for future
experiments.

With respect to spintronics application, magnetic sensors and non-volatile magnetic memories are conceivable. 
As compared to metal-based spintronics, metal-oxide structures are more versatile because of the ability to control 
the potential variation and spin polarization by external voltages \cite{Tian2012}.
Several spintronic experiments have been performed for 
{\em carbon} nanotubes using a two-terminal spin valve geometry \cite{Kim2002,Zhao2002}. 
This experimental setup makes it difficult, however, to separate spin transport from other effects, such as 
Hall effect, anisotropic magnetoresistance \cite{Jedema2001}, magneto-Coulomb \cite{vdMolen2006}
and interference effects \cite{Man2005}. A four-terminal non-local spin valve setup \cite{Jedema2001,Johnson1985,Jedema2002} 
with a single-wall nanotube can separate the spin current from the charge current \cite{Tombros2006}. Also, 
there are promising results for one-dimensional perovskite spintronic devices at a temperature lower than room temperature 
\cite{Li2016} (which is a key issue for commercial applications). Difficulties in device fabrication include 
the interface spin transport, and the positioning of nanowires or nanotubes with respect to the other components. 
The development of the system architecture may help overcome the difficulties associated with traditional electronics, 
and allow to develop one-dimensional spintronics technologies with good scalability, and lower power dissipation \cite{Tian2012}.

The details of the calculational approach are given in Section \ref{sec:computational}. In 
Section \ref{sec:stability}, we address the 
structural and magnetic properties. The main part of this study, the electronic structures of M-TNTs, 
is presented in Section \ref{sec:electronic}, followed by the optical and photocatalytic properties 
in Section \ref{sec:optical}. A brief summary is given in Section \ref{sec:summary}.

\section{Computational details}
\label{sec:computational}

All the calculations were carried out using plane-wave pseudopotentials in the Vienna ab initio simulation 
package \cite{vasp}. Spin-polarized calculations were employed to determine the structural, electronic, and 
magnetic properties. The generalized gradient approximation in the scheme of Perdew-Burke-Ernzerhof was used 
as an exchange-correlation functional for structure optimization. The plane-wave functions with cutoff energy 
550 eV and 1 $\times$ 1 $\times$ 3 $k$-mesh based on the Monkhorst-Pack method were utilized to obtain a 
converged force of 0.01 eV/{\AA}, and the total tolerance energy is 10$^{-6}$ eV. A periodic supercell 
along the tube axis (z axis) was considered with a vacuum distance of 20 {\AA} between nanotubes
in x and y directions to prevent the interaction between the neighboring TNTs. The convergence of the results 
was not affected when the parameters (cutoff energy, supercell size, and $k$-mesh) were increased.
The Heyd-Scuseria-Ernzerhof hybrid functional (HSE) \cite{hyb} was employed to calculate 
the formation energies, electronic structures, and optical properties.

The exchange-energy functional in the HSE scheme is written as a linear combination of short-range (SR) and 
long-range (LR) terms. The SR term is a PBE exchange functional ($E_{X}^{\mathrm{PBE,SR}}(\mu)$) mixed with a certain 
percentage of the exact exchange Hartree-Fock ($E_{X}^{\mathrm{HF,SR}}(\mu)$) contribution, while the LR term is defined 
by the PBE exchange functional ($E_{X}^{\mathrm{PBE,LR}}(\mu)$). The range-separation (screening) parameter $\mu$ 
is usually and also in this work chosen to be 0.2 {\AA}$^{-1}$. Therefore the exchange-correlation energy 
functional reads:
\begin{equation}
E_{x}^{\mathrm{HSE}} = a E_{x}^{\mathrm{HF,SR}}(\mu) + (1-a) E_{x}^{\mathrm{PBE,SR}}(\mu)
+ E_{x}^{\mathrm{PBE,LR}}(\mu),
\label{eq:hse}
\end{equation}
where $a$ is the called exchange mixing coefficient. The standard choice of $a$ in the HSE06 package 
is 25\%, but in order to reproduce the experimental TiO$_{2}$ bulk bandgap \cite{hybbul}, one has to 
choose a slightly smaller value, 22\%. However, both these $a$ values strongly overestimate the TNT bandgap.
In order to test the sensitivity of the bandgap with respect to variations of the mixing coefficient, we
have changed $a$ from 10\% to 28\%: the bandgap is found to be approximately given by 3.0 eV for 10\%,
and 4.3 eV for 28\%, with an almost linear increase in between; cf.\ Table \ref{0}.
The best $a$ value, reproducing the experimental 
TNT bandgap of 3.2 eV, is 14\%, which we have chosen in the following. Our results for the density of states
(not shown) also indicate that when increasing $a$ the conduction band shifts as a whole to higher energy (relative
to the Fermi energy), while the valence band stays rigid.

\begin{table*}[ht]
\caption{Titania nanotube bandgap versus mixing parameter $a$, as calculated with the HSE functional. 
The experimental value, 3.2 eV, is reproduced for $a = 14\%$. The low value of the optimal mixing parameter for TNT as
compared to bulk TiO$_{2}$ ($a = 22\%$) means that the electrons in the nanotube are less localized than those in
the bulk system, hence the TNT electrons are easily polarizable, implying good screening \cite{Marques2011}. This
also is an indication that there is no need for more sophisticated many-body technqiques like dynamical
mean-field theory.}
\centering
{
\begin{tabular}{|c|c|c|c|c|c|c|c|c|c|c|c|c|}
\hline
$a$ (\%) &10 & 12 & 14 & 16 & 18 & 20 & 22 & 25 & 28 \\\hline
Gap (eV) &3.0 & 3.1 & 3.2 & 3.4 & 3.6 & 3.7 & 3.8 & 4.0 & 4.3 \\\hline
\end{tabular}
}
\label{0}
\end{table*}

The dielectric function, $\varepsilon (\omega)=\varepsilon_{1}(\omega)+i\varepsilon_{2}(\omega)$, describes 
the optical response at the angular frequency $\omega$. First, $\varepsilon_{2}(\omega)$ is calculated on the
basis of the standard golden-rule expression, then $\varepsilon_{1}(\omega)$ is found by employing the
Kramers-Kronig relation. Finally, the absorption spectrum is determined by
\begin{equation}
\alpha(\omega) = \sqrt{2}\, \omega
\left( \sqrt{\varepsilon_{1}^{2}(\omega)+\varepsilon_{2}^{2}(\omega)}-\varepsilon_{1}(\omega) \right)^{1/2}.
\label{eq:absorption}
\end{equation}

\section{Stability of M-doped TNTs}
\label{sec:stability}

\begin{figure}
\begin{center}
\includegraphics[width=0.7\textwidth]{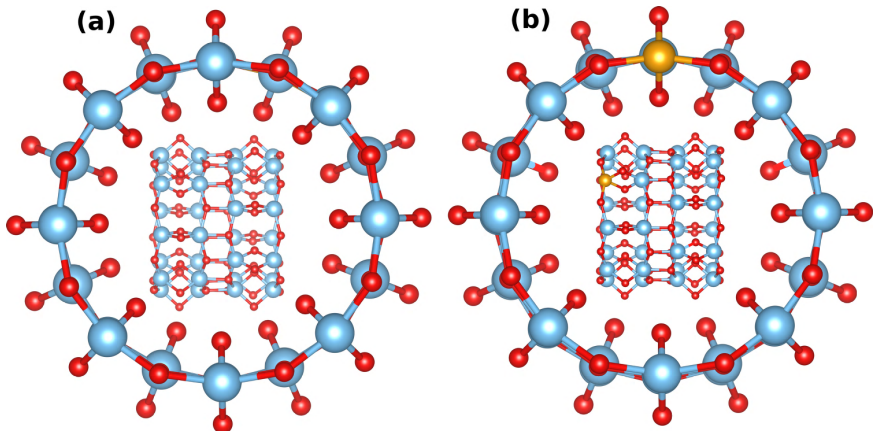}
\caption{Optimal configurations (top view) of pristine TNT (a) and metal-doped TNT (b), and their
side views (small figures). Red, sky-blue, and dark-yellow spheres represent O, Ti, and dopant atom, 
respectively.} 
\label{fig1}
\end{center}
\end{figure}

TiO$_{2}$ anatase nanotubes have been investigated experimentally and 
theoretically, see, e.g., \cite{st1,exp101} and \cite{AVB,Sz}, respectively, including details
of the geometries and their stabilities \cite{AVB,Sz,FADL,fadl2}. Figure \ref{fig1}(a) illustrates the 
pristine (8,0) TNT structure which contains 96 atoms along the tube length (10.49 {\AA}). 
The cation doped TNT, Fig.~\ref{fig1}(b), is created by replacing a Ti atom by the dopant. The bond length 
between dopant and Ti atom elongates as the ionic radius of the dopant atom increases for all dopants, as 
given in Tables \ref{1}, \ref{2}, and \ref{3}. The stability of the metal-doped TNT (M-TNT) is determined
from the formation energy ($E^{f}$):
\begin{equation}
E^{f} = E^{\mathrm{M-TNT}}+\mu^{\mathrm{Ti}}-(E^{\mathrm{TNT}}+\mu^{\mathrm{M}}),
\label{eq:formation}
\end{equation}
where $E^{\mathrm{M-TNT}}$ and $E^{\mathrm{TNT}}$ denote the total energies of metal-doped and pristine TNT, 
respectively; $\mu^{\mathrm{Ti}}$ and $\mu^{\mathrm{M}}$ are the chemical potentials of Ti and the dopant atom, 
the latter assumed to be given by the energy of the isolated metal atom.

\begin{table*}[ht]
\caption{Ionic radius (\AA) (cf.~\cite{tab1}), electronegativity (cf.~\cite{tab2}), bond lengths (\AA), formation 
energies ($E_{f}$ (eV)), magnetic moments ($\mu_{B}$), and bandgap (eV) of selected 3$d$-metal doped TNTs. The
pristine system (Ti column) is included for easy reference. The last three rows indicate whether the
respective system is useful for spintronic, optical, and photocatalytic applications: the \ding{51} means ``clear
improvement compared to pristine TNT'', and the \ding{55} ``no improvement'', w.r.t.\ that application.}
\centering
{
\begin{tabular}{|c|c|c|c|c|c|c|c|c|c|c|c|c|}
\hline
$X$ &Sc& Ti& V& Cr& Mn&Fe&Co&Ni&Cu&Zn\\\hline

Ionic radius &0.75&0.61&0.54&0.44&0.46&0.78&0.55&0.48&0.54&0.74\\\hline 

Electronegativity &1.36&1.54&1.63& 1.66& 1.55&1.83& 1.88& 1.91& 1.90& 1.65\\\hline 
 
O-$X$ bond &2.05&1.90 & 1.89 & 1.83 & 1.87 & 1.88 & 1.85 & 1.87 & 1.96 & 2.00 \\\hline  

$E_{f}$ &2.91 & -- & 1.55 & 2.59 & 3.71 & 7.18 & 7.61&9.6 & 11.90 & 13.16 \\\hline

Magnetic moment &1.0 &0.0 & 1.0& 2.0 & 3.0 & 4.0 & 1.0 & 0.0 & 1.0 & 2.0 \\\hline 

Bandgap &2.8 &3.2 & 2.8& 2.0 & 2.5 & 1.7 & 1.5 & 2.0 & 1.7 & 2.4 \\\hline

Spintronic &\ding{55}&--&\ding{51}&\ding{51} &\ding{55}&\ding{51} &\ding{51} &\ding{55}& \ding{55} & \ding{55}\\\hline

Optical &\ding{51}&--&\ding{51}&\ding{51}& \ding{51}&\ding{51} &\ding{51}&\ding{51}& \ding{51} & \ding{51}\\\hline

Photocatalytic &\ding{55}&--&\ding{55}&\ding{51}&\ding{51}&\ding{51}&\ding{51}&\ding{55}&\ding{55} & \ding{55}\\\hline
\end{tabular}
}
\label{1}
\end{table*}

We find the formation energy to be roughly proportional to the number of electrons in the dopant atom, and related 
to the ionic radius of the larger ions (Sc, Y, La, and Zr). The most stable M-TNT is Ta-TNT because the ionic radius 
and electronegativity of Ta are very close to the corresponding values of the Ti ion. The formation energies of 4$d$-TNTs 
are higher than the corresponding ones in the same group of 3$d$- and $5d$-TNTs, which can be attributed to the 
large ionic size of 4$d$ dopants as compared to the ionic size of 3$d$ dopants group, and the low electronegativity of 
4$d$ dopants as compared to the electronegativity of 5$d$ dopants (Tables \ref{1}, \ref{2}, and \ref{3}). 

Regarding the magnetic moment, the difference between the number of outer shell electrons in the metal dopant and the Ti 
atom determines the magnetic properties of M-TNTs as is apparent, e.g., for Sc- to Fe-TNTs. For the nearly full outer 
shell Co, the coupling between the outer shell electrons can explain the magnetic moment 
of Co-TNT: two outer shell electrons are coupling and the third one is unpaired (low spin state), 
hence the net magnetic moment is 1 $\mu_{B}$ to a good approximation. 

On the other hand, for Co one needs one electron to fill the outer shell, thereby a hole will be created in Co-TNT,
and the magnetic moment becomes also 1 $\mu_{B}$. For the full-filled outer shell atoms (Ni, Cu, and Zn), the magnetic moments
are 0 for Ni due to the closed (inert) shell, 1 $\mu_{B}$ for Cu (the oxidation number is (+3)) and a hole is created, 
and 2 $\mu_{B}$ for Zn (the oxidation number is (+2)) and two holes are created; see Table \ref{1}. The same trend appears 
in the 4$d$-TNTs (Table \ref{2}) except for Tc- and Ru-TNT. As compared to the same atom group (Mn), the magnetic 
moment for Tc-TNT may be attributed to the low spin state because the coupled states are the same as in Co-TNT. For the Ru
doped structure, the magnetic moment is 2 $\mu_{B}$: two electrons of the outer shell are coupled, the other two are 
uncoupled (low spin state). From another point of view, 
two electrons are needed to close the outer shell of Ru so two holes are created in the structure, 
hence the magnetic moment becomes 2 $\mu_{B}$.
For the selected 5$d$ dopants, the magnetic moments can be explained by comparison with the above
discussed atoms in the {\em same group} of the periodic table, hence with the same number of 
outer shell eletrons. For example, the Sc, Y, and La dopants (group 3B) have the same magnetic moment 
of 1 $\mu_{B}$ (Table \ref{3}).

\begin{table*}[htb]
\caption{Ionic radius (\AA) (cf.~\cite{tab1}), electronegativity (cf.~\cite{tab2}), bond lengths (\AA), formation 
energies ($E_{f}$ (eV)), magnetic moments ($\mu_{B}$), and bandgap (eV) of selected 4$d$-metal doped TNTs. 
The last three rows indicate whether the
respective system is useful for spintronic, optical, and photocatalytic applications: the \ding{51} means ``clear
improvement compared to pristine TNT'', and the \ding{55} ``no improvement'', w.r.t.\ that application.}
\centering
{
\begin{tabular}{|c|c|c|c|c|c|c|c|c|c|c|c|c|}
\hline
$X$  &Y&Zr&Nb &Mo &Tc &Ru&Rh&Pd &Ag &Cd\\\hline

Ionic radius &0.90&0.72&0.64&059&0.60&0.68&0.67& 0.62&0.75 &0.95\\\hline 
 
Electronegativity &1.22&1.33&1.6& 2.16&1.90& 2.20& 2.28& 2.20& 1.93& 1.69\\\hline
 
O-$X$ bond &2.22&2.06 & 1.96 & 1.97 & 1.96 & 1.96 & 1.97 & 1.99 & 2.11 & 2.26 \\\hline  

$E_{f}$ &2.90 &$-1.10$&1.03 & 4.06 & 5.47 & 6.79 & 8.54 & 11.62 & 14.05&14.20 \\\hline

Magnetic moment &1.0 &0.0 & 1.0 & 2.0 & 1.0 & 2.0 & 1.0 & 0.0 & 1.0 &2.0 \\\hline 

Bandgap &2.8 &3.2 & 2.9& 2.4 & 2.5 & 1.6 & 1.2 & 2.2 & 1.8 & 2.4 \\\hline

Spintronic     &\ding{55}&\ding{55}&\ding{51}&\ding{51} &\ding{51}&\ding{51} &\ding{51} &\ding{55}& \ding{55} &\ding{55}\\\hline

Optical        &\ding{51}&\ding{55}&\ding{51}&\ding{51} & \ding{51}&\ding{51}&\ding{51}&\ding{51}& \ding{51}& \ding{51}\\\hline

Photocatalytic &\ding{55}&\ding{55}&\ding{51}&\ding{52}& \ding{55}&\ding{51}&\ding{55}&\ding{55}& \ding{55} & \ding{55}\\\hline
\end{tabular}
}
\label{2}
\end{table*}

\begin{table*}[htb]
\caption{Ionic radius (\AA) (cf.~\cite{tab1}), electronegativity (cf.~\cite{tab2}), bond lengths (\AA), formation 
energies ($E_{f}$ (eV)), magnetic moments ($\mu_{B}$), and bandgap (eV) of selected 5$d$-metal doped TNTs. 
The last three rows indicate whether the
respective system is useful for spintronic, optical, and photocatalytic applications: the \ding{51} means ``clear
improvement compared to pristine TNT'', and the \ding{55} ``no improvement'', w.r.t.\ that application.}
\centering
{
\begin{tabular}{|c|c|c|c|c|c|}
\hline
$X$  &La&Ta &W &Pt &Au\\\hline
 
Ionic radius &1.03&0.64& 0.60& 0.63&0.75\\\hline 
 
Electronegativity &1.1&1.50&2.36& 2.28& 2.54\\\hline 
 
O-$X$ bond &2.36& 1.92 & 1.90 & 2.02 & 2.2 \\\hline  

$E_{f}$ &2.59 &$-$1.56 & 1.94&9.08 & 13.21  \\\hline

Magnetic moment &1.0 &1.0 & 2.0 & 0.0 & 1.0 \\\hline 

Bandgap &2.8 &3.0 & 2.4 & 1.8 & 2.2 \\\hline 

Spintronic &\ding{55}&\ding{51} &\ding{51}&\ding{51} &\ding{55}\\\hline

Optical &\ding{51}&\ding{51}&\ding{51}&\ding{51}& \ding{51}\\\hline

Photocatalytic &\ding{55}&\ding{51}&\ding{51}&\ding{55} &\ding{55}\\\hline
\end{tabular}
}
\label{3}
\end{table*}

\section{Electronic structure}
\label{sec:electronic}

\begin{figure}
\begin{center}
\includegraphics[width=0.8\textwidth]{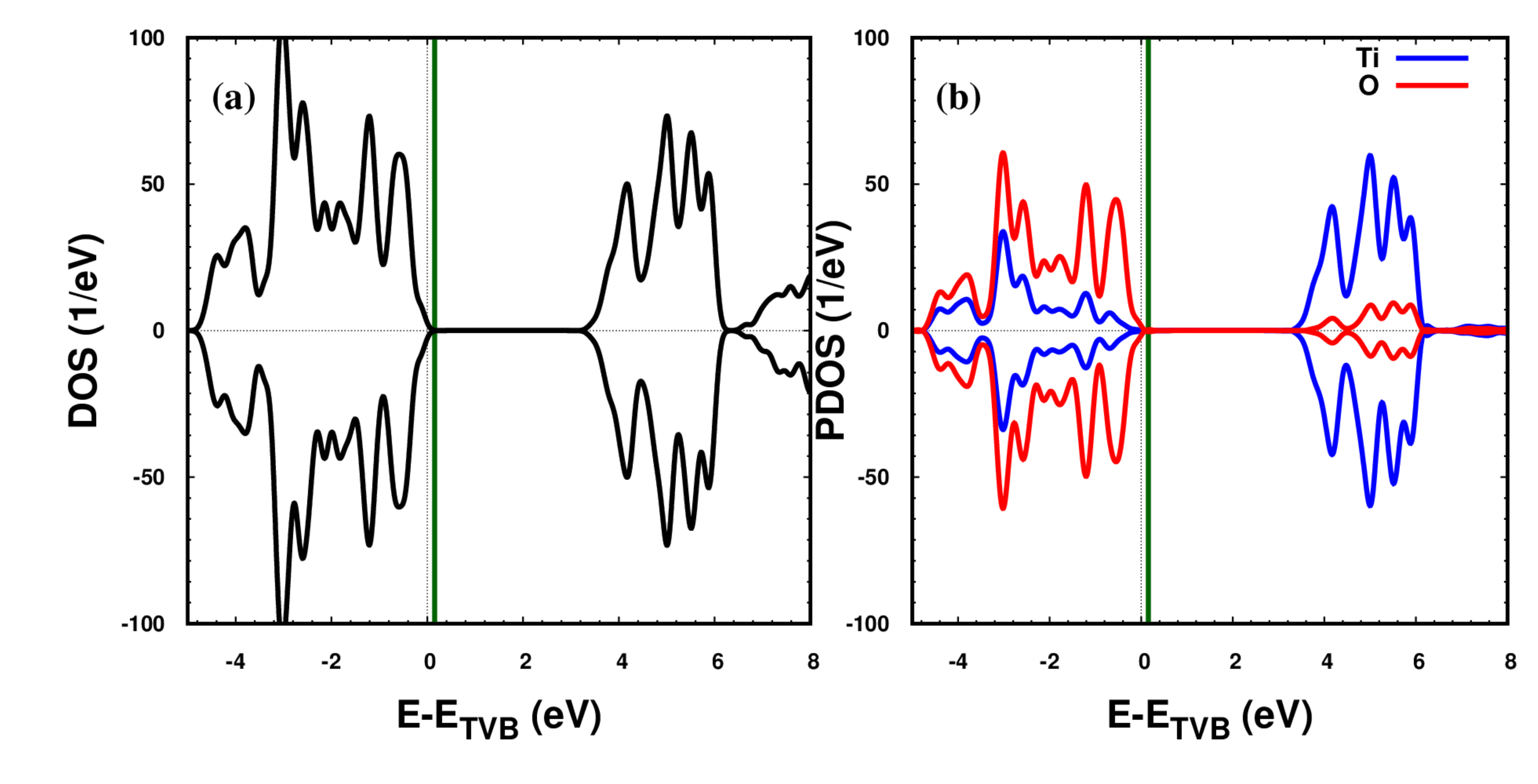}
\caption{Density of states (DOS) (a), and partial density of states (PDOS) (b), of pristine TNT. The energy
is given relative to the top of the valence band (TVB), and the green vertical line indicates the Fermi energy.
The overall features of the DOS of pristine TNT, obtained here within {\em hybrid} DFT, agree well with the
result found previously \cite{fadl2} on the basis of DFT-GGA, provided the ``scissors'' operation is applied; see
\cite{fadl2} and references therein.} 
\label{fig2}
\end{center}
\end{figure}

The density of states (DOS) and projected density of states (PDOS) for pristine TNT are shown in 
Figs.~\ref{fig2}(a,b). The bandgap of 3.20 eV is in good agreement with the experimental bandgap, 
3.18--3.23 eV \cite{tnte1,tnte2} (since we have chosen the mixing parameter, $a$, accordingly, 
cf.\ Section \ref{sec:computational}). The O states are dominant in the valence band (VB), 
while the Ti states dominate in the conduction band (CB) (Fig.~\ref{fig2}(b)). 

\begin{figure}
\begin{center}
\includegraphics[width=0.6\textwidth]{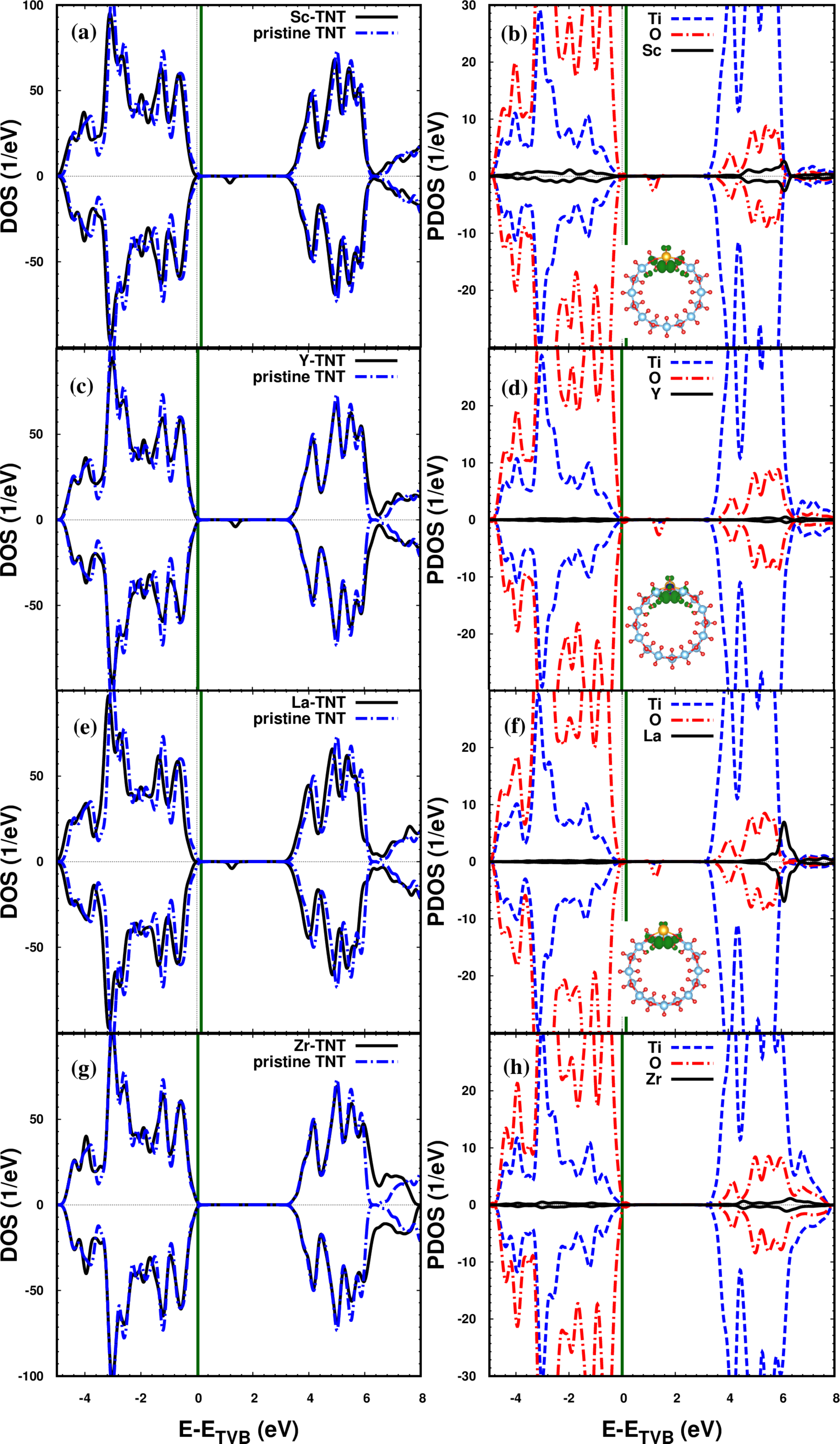}\\
\caption{Density of states (DOS) and partial density of states (PDOS) for Sc- (a,b), Y- (c,d), La- (e,f), and Zr-doped (g,h)
TNTs. The energy is given relative to the top of the valence band (TVB), and the green vertical line indicates the Fermi energy. 
The inset figures show the corresponding spin density isosurfaces.} 
\label{fig3}
\end{center}
\end{figure}

In order to structure the presentation, we discuss the electronic structures according to the groups 
in the periodic table. For 3B group dopants (Sc, Y, La), due to their oxidation number of (+3) as compared 
to (+4) for Ti, the DOS is asymmetric between the two spin components, and intermediate states are created at 
1.1, 0.9, and 1.3 eV for Sc-, Y-, and La-TNTs, respectively. The dopants introduce a hole in the structure. 
The intermediate states decrease the bandgap to 2.8 eV. As compared to the pristine case, 
the VB edge, the Fermi energy, and the CB edges remain unchanged, see Figs.~\ref{fig3}(a,c,e). 
The intermediate states are dominated by states from O atoms which are close to the dopant atom; 
see the PDOSs of the doped structures (Figs.~\ref{fig3}(b,d,f)) and their spin density isosurfaces
(respective insets). Note that the reduction in the bandgap may enhance the spectral activity, 
even though the photocatalytic properties are not being improved due to the location of the intermediate 
states. In fact, the intermediate states in these doped structures represent recombination centers.
Experimentally, Y-TNT was successfully synthesized by a microwave refluxing method, and 
it was found that the optical activity is increased \cite{Y0} which indicates a decrease in the bandgap,
consistent with our findings. 

For the 4B group dopant considered (Zr), the DOS of Zr-TNT has the same DOS as pristine TNT 
due to the similarity of the Zr and Ti outer shells, so there is no change in the bandgap. 
The PDOS shows there is no contribution of Zr states in the bandgap and at the band 
edges (Figs.~\ref{fig3}(g,h)). 

\begin{figure}
\begin{center}
\includegraphics[width=0.6\textwidth]{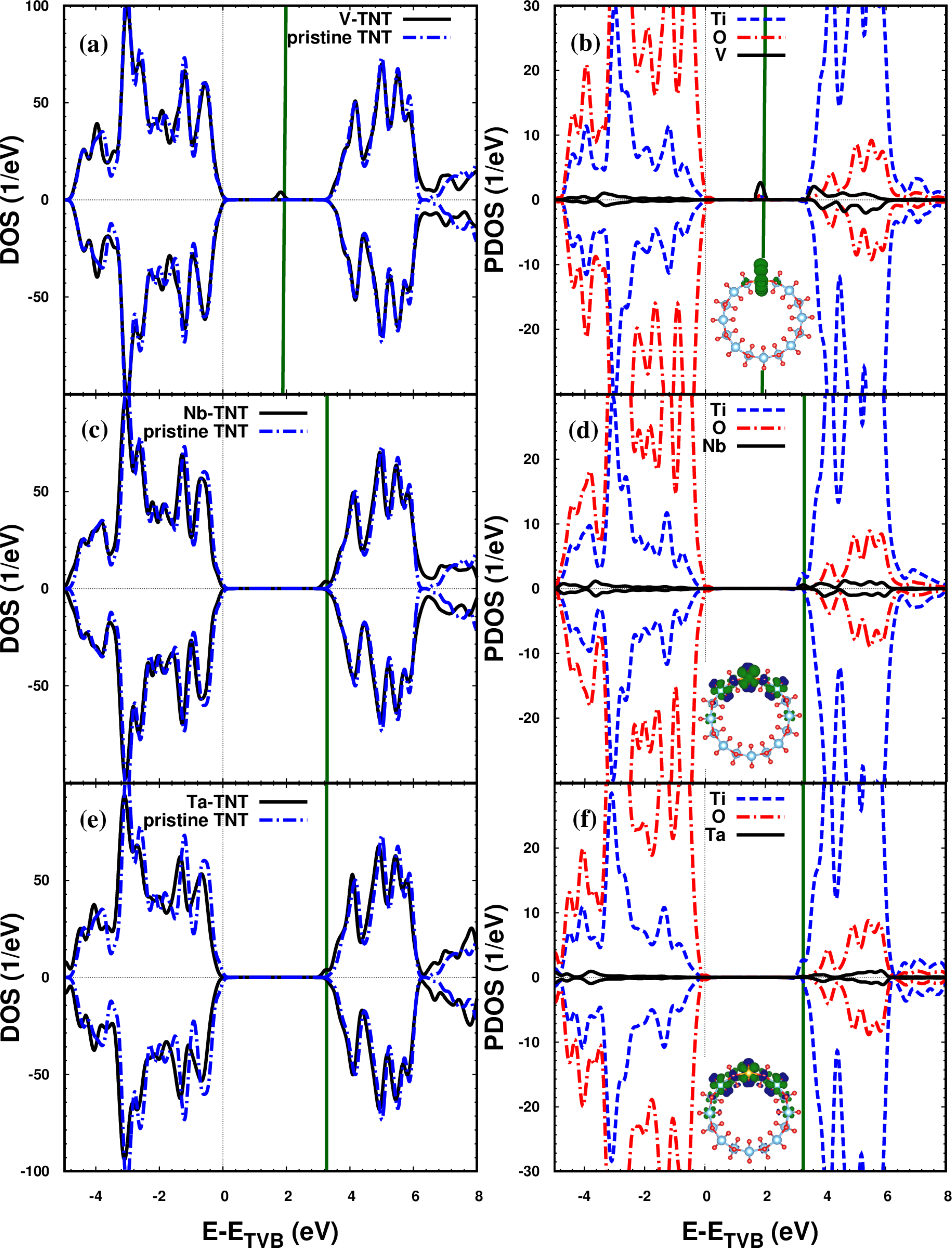}\\
\caption{Density of states (DOS) and partial density of states (PDOS) for V- (a,b), Nb- (c,d), and Ta-doped (e,f)
TNTs. The energy is given relative to the top of the valence band (TVB), and the green vertical line indicates the Fermi energy. 
The inset figures show the corresponding spin density isosurfaces.} 
\label{fig4}
\end{center}
\end{figure}

Since the dopants of the 5B group (V, Nb, Ta) have an additional electron compared to Ti, an asymmetric 
behavior of the two spin components of the DOS arises. The additional states in the bandgap are located at 1.9 eV for V-TNT, 
and at the edge of the CB for Nb- and Ta-TNTs (Figs.~\ref{fig4}(a,c,e)). The bandgaps are 2.8, 2.9, and 3.0 eV 
for V-, Nb-, and Ta-dopants, respectively. The contributions of V (dopant) states to the midgap states are significant 
in V-TNT, while for Nb- and Ta-TNTs the generated states derive from the Ti (host) states, see Figs.~\ref{fig4}(b,d,f). 
The V dopant increases the optical activity of pristine TNT without any improvement in the photocatalytic properties 
due to the position of the midgap states. The generated states at the bottom of the CB, 
due to Nb and Ta, slightly extend the optical absorption range, enhance the photocatalytic efficiency, 
and increase the conductivity. Experimental results indeed have shown that V and Nb dopants decrease 
the bandgap of pristine TNT \cite{Lu2014}, and of thin films \cite{Nb2}. It is worth mentioning that the 
reduced bandgap of TNT due to doping with Nb is very similar to the reduced bandgap of TiO$_{2}$ thin films with 
the same dopant, namely 0.3 eV. Furthermore, the experiments observed that Nb-TNT is an n-type 
conductor \cite{nanoimp}, consistent with our results. Also, it was found experimentally
that Nb \cite{Nb1} and Ta \cite{TaNT} dopings enhance the photocatalytic activity of TNT for water splitting. 
In addition, ferromagnetic behavior was reported for V-TNT \cite{Vmag}. Clearly, due to the 
metallic spin-up states of Nb- and Ta-TNTs at the Fermi energy, 
these systems may be useful for spintronic applications. 

\begin{figure}
\begin{center}
\includegraphics[width=0.6\textwidth]{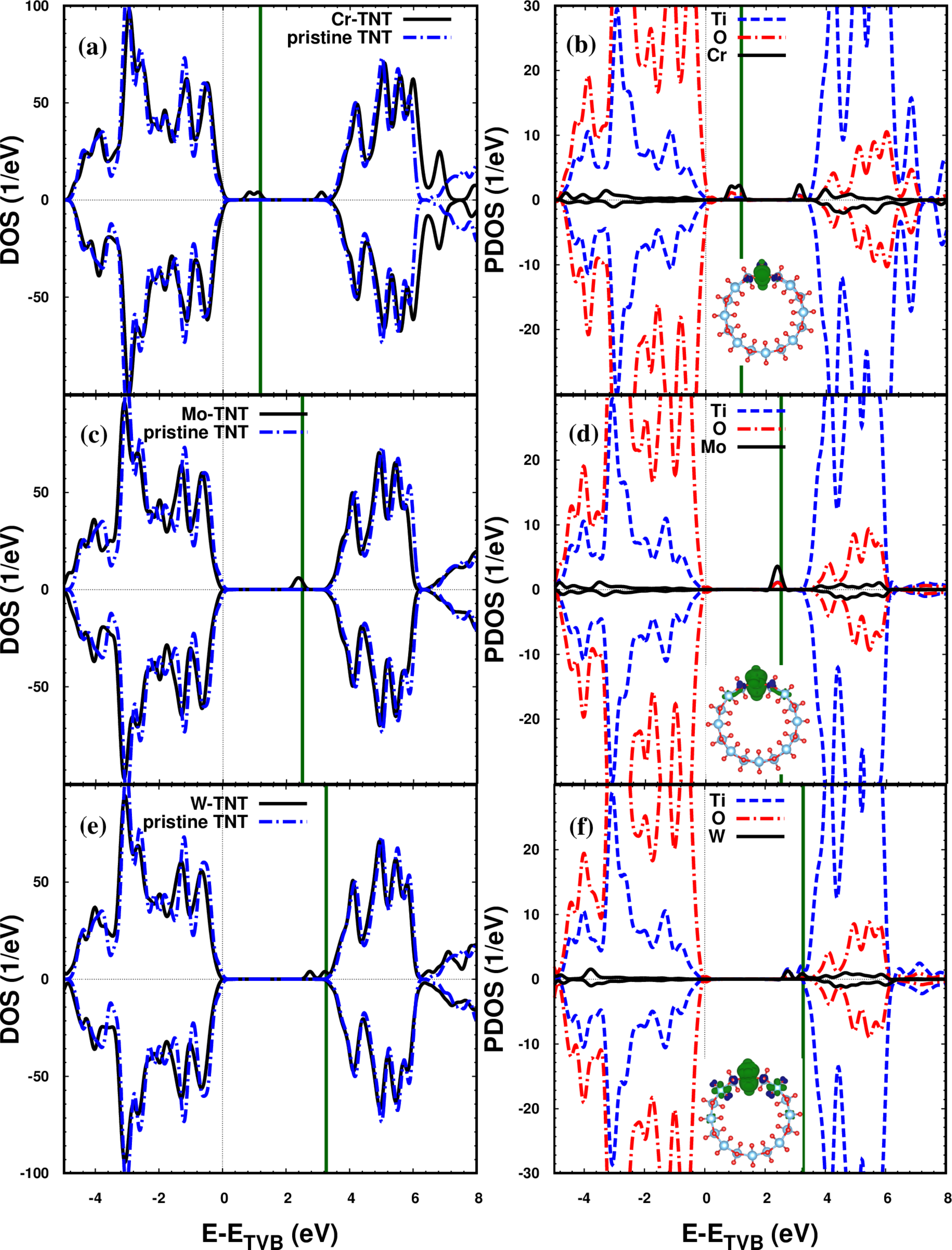}\\
\caption{Density of states (DOS) and partial density of states (PDOS) for Cr- (a,b), Mo- (c,d), and W-doped (e,f) TNTs. 
The energy is given relative to the top of the valence band (TVB), and the green vertical line indicates the Fermi 
energy. The inset figures show the corresponding spin density isosurfaces.} 
\label{fig5}
\end{center}
\end{figure}

For the next group, 6B (Cr, Mo, W), there are two additional electrons from the dopants which again implies 
an asymmetry in the spin resolved DOS (Figs.~\ref{fig5}(a,c,e)). The extra states in the bandgap are 
created in the range of 0.5--1.3 eV, and below the CB edge for Cr-TNT. The generated states appear at 2.7 eV for Mo-TNT, 
and below the CB edge for W-TNT. The gaps of Cr-, Mo-, and W-TNTs are 2.0, 2.4, and 2.4 eV, respectively.
The doping with Cr and W improves 
the optical and photocatalytic activities under visible light irradiation, due to the positions of the states generated 
within their gaps. For the Cr dopant, the created states are close to the VB and below the CB, therefore the probability 
of creating a trapping center is small \cite{nano}. The locations of the intermediate states for the Mo dopant 
increases the optical activity as compared to the pristine case. Experimentally, an enhancement of the photocatalytic 
activity for Cr-TNT \cite{Cr4d} as well as a gap
narrowing for Mo-TNT \cite{Mo0} were measured. For W doping, a reduction of the  
gap of nanoparticles was experimentally found \cite{Wnp0}. The created bandgap states are related 
to the dopant states (Figs.~\ref{fig5}(b,d,f)), and the Fermi energy crosses them for the spin up component, 
therefore Cr-, Mo-, and W-TNTs can be beneficial for spintronic applications. A ferromagnetic behavior at room 
temperature was observed for Cr-TNT \cite{CerFer}. 

\begin{figure}
\begin{center}
\includegraphics[width=0.6\textwidth]{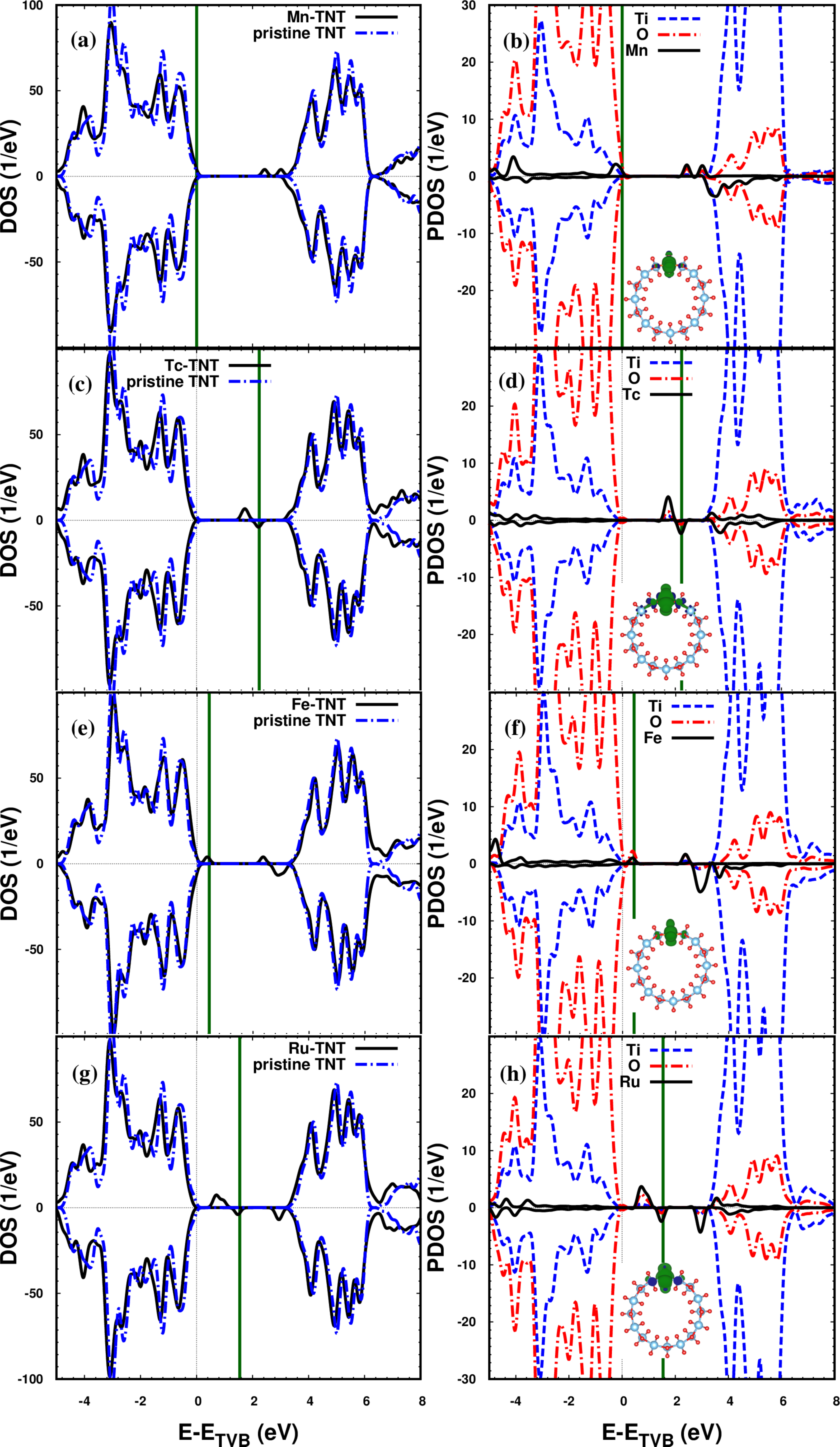}\\
\caption{Density of states (DOS) and partial density of states (PDOS) for Mn- (a,b), Tc- (c,d), Fe- (e,f), and 
Ru-doped (g,h) TNTs. The energy is given relative to the top of the valence 
band (TVB), and the green vertical line indicates the Fermi energy. The inset figures show the corresponding 
spin density isosurfaces.} 
\label{fig6}
\end{center}
\end{figure}

Considering the dopants in the 7B group, Mn and Tc (Figs.~\ref{fig6}(a,c)), the excess single electron in the 
doped structures leads again to a spin asymmetry in the DOS,
and also decreases the bandgap. The created states which appear below the CB edge and at 2 eV for Mn-TNT 
and Tc-TNT, respectively, reduce the bandgap to 2.4 eV for both dopants. Hence Mn and Tc dopings improve the optical
properties. The Mn dopant can also enhance the photocatalytic activities due to the location of the created states. 
This result confirms the experimental observations of the photocatalytic performance for Mn-TNT \cite{Mn2}. 
Figures \ref{fig6}(b,d) show the PDOS of Mn- and Tc-doped TNT; in both cases, the 
dopant states have their main contributions in the generated states. For Tc-TNT, 
the spin down midgap states are created at the Fermi energy, so Tc-TNT can be used for spintronic applications. 
Recently, Tc doped bulk TiO$_{2}$ was synthesized \cite{tc}.

\begin{figure}
\begin{center}
\includegraphics[width=0.6\textwidth]{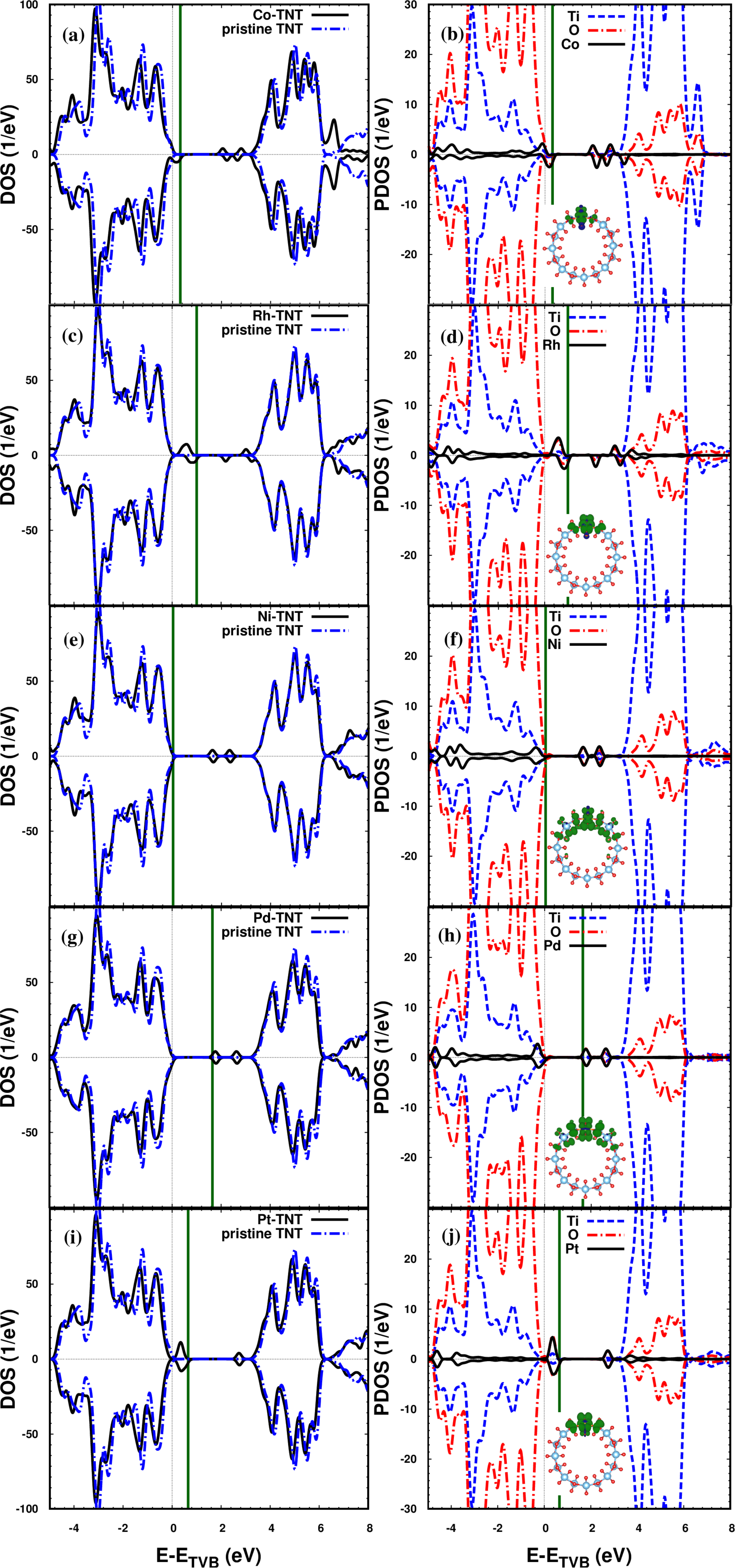}\\
\caption{Density of states (DOS) and partial density of states (PDOS) for Co- (a,b), Rh- (c,d), Ni- (e,f), Pd- (g,h), 
and Pt-doped (g,h) TNTs. The energy is given relative to the top of the valence 
band (TVB), and the green vertical line indicates the Fermi energy. The inset figures show to the corresponding 
spin density isosurfaces.} 
\label{fig7}
\end{center}
\end{figure}

Regarding the 8B group, first two columns (Fe, Ru, Co, Rh), these atoms have nearly full outer shells. 
As compared to the previous dopants, the created states are more spread inside the bandgap, and appear
near the edges of the VB and the CB as well, which implies a stronger reduction of the bandgap (Figs.~\ref{fig6}(e,g) 
and Figs.~\ref{fig7}(a,c)). The bandgaps are 1.7, 1.6, 1.5, and 1.2 eV for Fe-, Ru-, Co-, and Rh-TNTs, 
respectively. Except for the Rh dopant, all of them will be good candidates 
for enhancing the optical and photocatalytic activities of TNT. For Fe doping, an improvement of the
photocatalytic activity was found experimentally \cite{Fed2,nanoimp}. Also Ru-TNT  
showed a reduced bandgap as compared to TNT in an experiment \cite{Ru00}. 
Strong contributions of dopant states exist in the midgap states at the band edges,
as shown in the PDOS and the corresponding isosurfaces spin (Figs.~\ref{fig6}(f,h) and 
Figs.~\ref{fig7}(b,d)). Furthermore, the Fermi energy is located inside a portion of 
the created states for one spin component, therefore such doped TNTs can be used in 
spintronic applications. We note that ferromagnetic behavior was experimentally 
observed for Co-TNT \cite {spintCo}.

We turn now to the closed outer shell atoms, namely the last column in the 8B group (Ni, Pd, Pt). 
The DOS of Ni- and Pd-TNTs are very similar; the intermediate states of Ni-TNT 
are located near the middle of bandgap. On the other hand, for Pt-TNT they are located at 
the edges of the VB and the CB (Figs.~\ref{fig7}(e,g,i)). Due to the inert outer shell and 
zero magnetic moment, the DOS shows a spin-symmetric behavior.
The gaps of those structures are 2.0, 2.2, and 1.8 eV for Ni-, Pd-, and Pt-TNTs, respectively, 
all of them being smaller than the pristine TNT gap.
Figures \ref{fig7}(f,h,j) show 
the contributions of the Ni, Pd, and Pt states to the midgap states. Clearly, this group of dopants can 
improve the optical activity of TNT. But only Pd doping can enhance the photocatalytic activity of TNT 
due to the contribution of Pd states at the VB and CB edges. The decrease in the gap of Ni-TNT, as well as
of Pd- and Pt-doped nanoparticles, was experimentally reported \cite{Ni1d,Pdnp0,Ptnp0}.

\begin{figure}
\begin{center}
\includegraphics[width=0.6\textwidth]{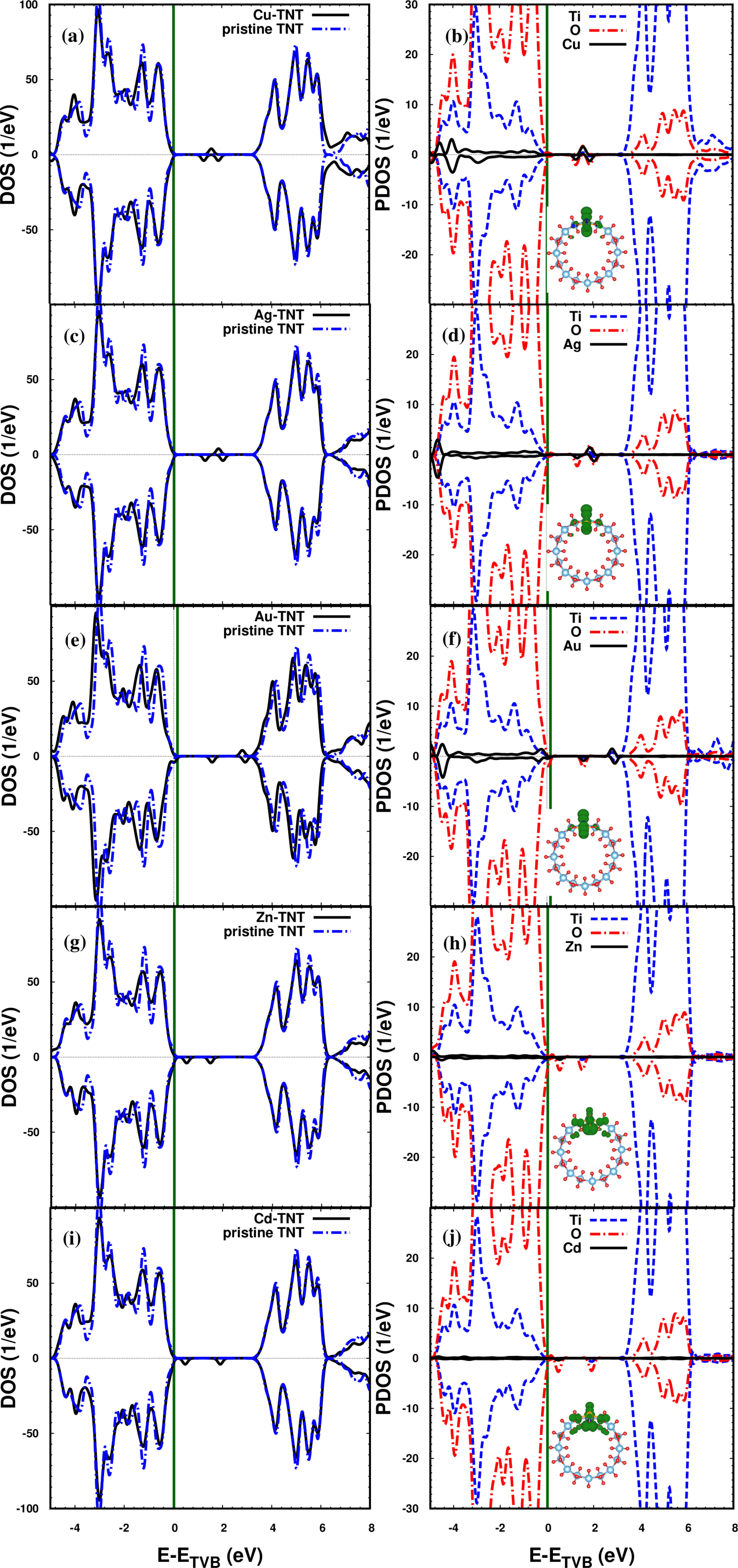}\\
\caption{Density of states (DOS) and partial density of states (PDOS) for Cu- (a,b), Ag- (c,d), Au- (e,f), 
Zn- (g,h), and Cd-doped (i,j) TNTs. The energy is given relative to the top of the valence 
band (TVB), and the green vertical line indicates the Fermi energy. The inset figures show the 
corresponding spin density isosurfaces.} 
\label{fig8}
\end{center}
\end{figure}
   
The next closed outer shell group is 1B (Cu, Ag, Au). The effect of Cu and Ag on the DOS is seen
in the midgap states of Cu- and Ag-TNTs, and in the midgap and CB edge states for Au-TNT (Figs.~\ref{fig8}(a,c,e)). 
The bandgap becomes 2.1, 1.8, and 2.2 eV for Cu-, Ag, and Au-TNTs, respectively. Figures \ref{fig8}(b,d,f) show 
also that the contributions from O atoms surrounding the dopant atoms are stronger than the 
contributions of the metal dopant states. Although the Fermi energy 
of Au-TNT crosses the created state, it is not suitable for spintronic applications because the 
contribution of the O states is dominant. 
Reduced bandgaps, compared to pristine TNT, have been observed experimentally for 
Cu- and Ag-TNTs, and for Au doped TiO$_{2}$ nanoparticles \cite{nanoimp,Agn0,Ptnp0}.
This dopant group hence can only improve the optical activity. 

The last closed outer shell group is 2B (Zn, Cd). The effect of both dopants is the same in 
the DOS (Figs.~\ref{fig8}(g,i)), with a small difference, however, in the location of the 
created midgap states. Figures \ref{fig8}(h,j) show that the midgap states are dominated by
contributions from O states.
Due to the reduction in the bandgap (2.2 eV for both dopants), Zn and Cd doping can improve the 
optical activity of pristine TNT, in agreement with experimental results for Zn-TNT \cite{Zn00}, 
and for Cd-doped TiO$_{2}$ nanoparticles \cite{Cdnp}.
The bandgaps and the potential applications (optical, photocatalytic activities, and spintronics) 
are summarized in Tables \ref{1}, \ref{2}, and \ref{3}.

\section{Optical properties and clean fuel production}
\label{sec:optical}

\begin{figure}
\begin{center}
\includegraphics[width=0.7\textwidth]{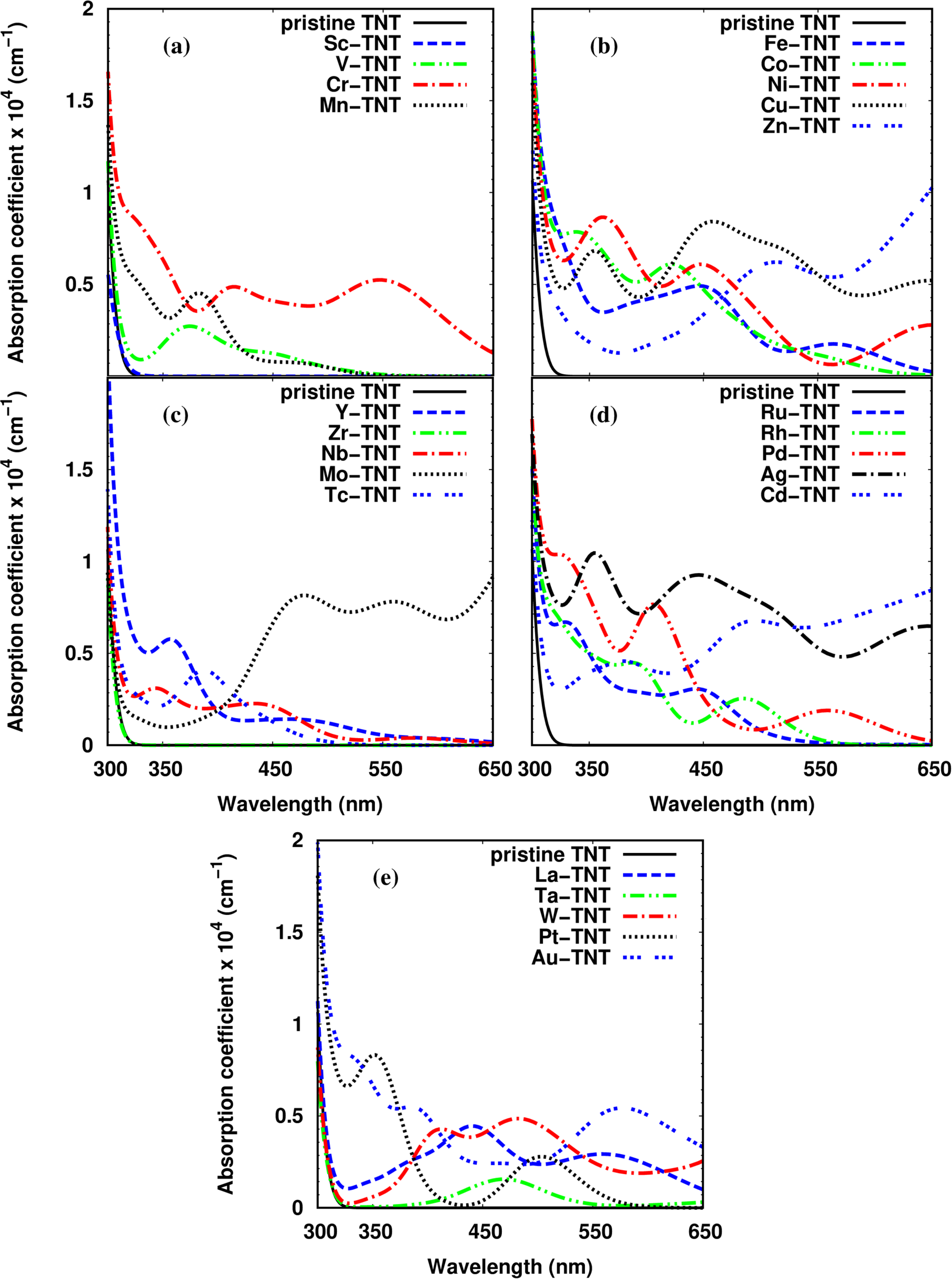}
\caption{Absorption coefficients for 3$d$- (a,b), 4$d$- (c,d), and selected 5$d$-doped (e) TNTs.} 
\label{fig9}
\end{center}
\end{figure}

\begin{figure}
\begin{center}
\includegraphics[width=0.85\textwidth]{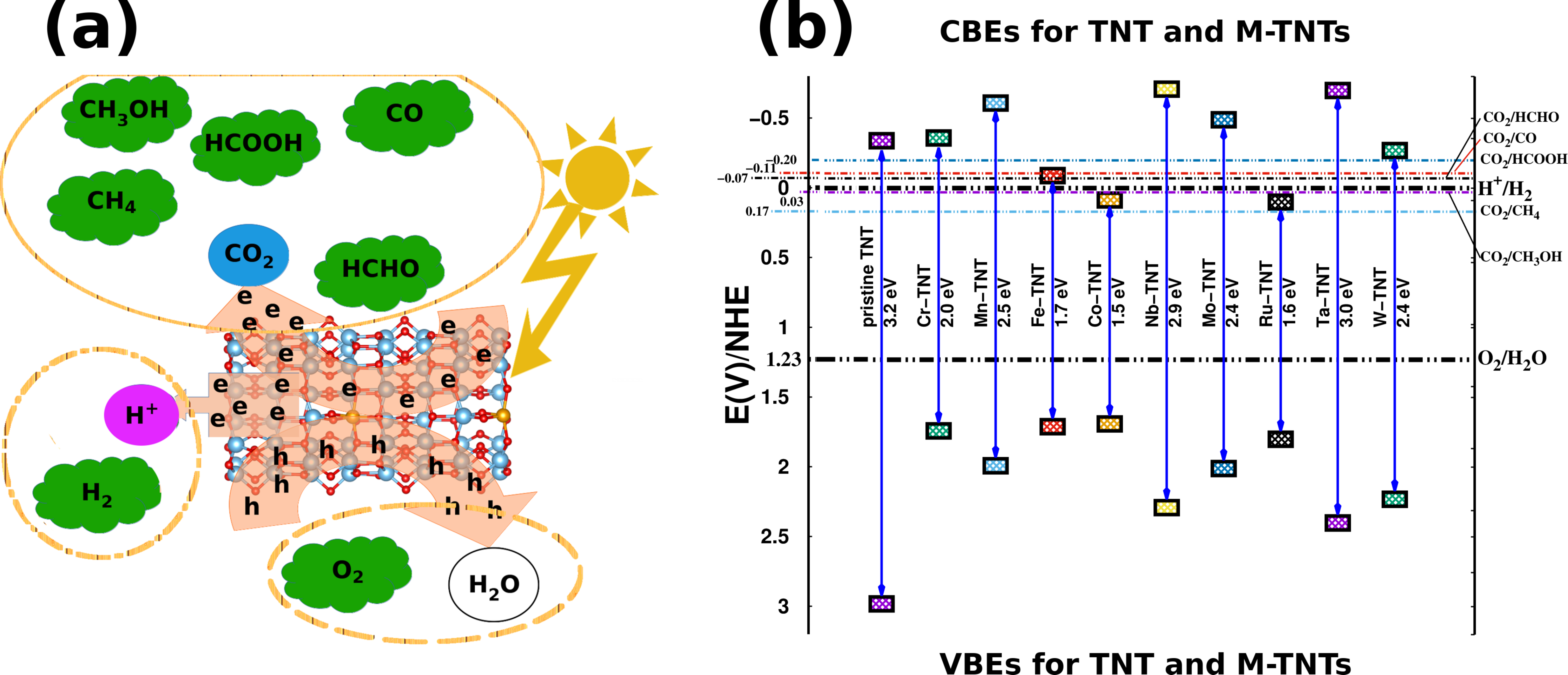}
\caption{Fundamentals of pristine and M-doped TNTs for photocatalytic water 
splitting and carbon dioxide reduction for clean fuel production.
(a) Schematic overview of concepts and chemical reactions. 
(b) Summary of band edge positions of the relevant systems. The dashed horizontal lines indicate 
the water redox and carbon dioxide potentials at pH = 0. The values are given with respect to the NHE 
(normal hydrogen electrode) potential (in Volts).} 
\label{fig10}
\end{center}
\end{figure}

Figure \ref{fig9} summarizes the absorption coefficients 
of pristine, 3$d$-, 4$d$-, and selected 5$d$-TNTs using Eq.~(\ref{eq:absorption}). All doped structures show an
increase of light absorption, in particular, extending to a wider wavelength range, as compared to the pristine 
case, except for Sc and Zr. The absorption edge of most M-TNTs is shifted towards lower energy (redshift). Also, 
additional absorption peaks are observed for M-TNTs in the low energy range. The absorption edge is not only related to 
the bandgap but also depends on the number of electrons of the dopant: naturally, as the electron number 
increases the light absorption and its range increase (see, e.g., Fig.~\ref{fig9}(b) as 
compared to Fig.~\ref{fig9}(a)).

In Fig.~\ref{fig10}(a) we display a schematic figure to elucidate the basic idea of H$_{2}$O splitting and 
CO$_{2}$ reduction using M-TNTs as photocatalysts. When M-TNTs absorb light with energy larger than or equal 
to the bandgap of the photocatalysts (M-TNTs), the electron-hole pairs are separated and migrate to the surface 
of the photocatalyst, electrons are excited near the CB edge (CBE) and holes are created near the 
VB edge (VBE). As the bandgap of 
the photocatalyst decreases, the range of light absorption by M-TNTs increases. During the migration of the
photogenerated charges, they may recombine for many reasons, and the number of them will be reduced in the 
photocatalytic H$_{2}$O splitting and CO$_{2}$ reduction. The midgap states 
can be considered as active recombination (trapping) centers for photogenerated charges, 
so the midgap states are not beneficial. 

After the arrival of the photogenerated charge pairs at the M-TNT surface, the photogenerated electrons will reduce
the adsorbed H$_{2}$O on M-TNT to form H$_{2}$ fuel gas, and the photogenerated holes will oxidize H$_{2}$O to 
form O$_{2}$ on different active surface sites. Also, the photogenerated electrons can be used to reduce the 
adsorbed CO$_{2}$ on the M-TNTs to several natural fuels (Fig.~\ref{fig10}(a)). 
The two previous processes can only occur when the CBE is more negative than the H$^{+}$ 
(protons which were produced from the water oxidation process) or CO$_{2}$ reduction potentials,
and the VBE is more positive than the H$_{2}$O oxidation potential. 
The natural fuel type which can be gained depends on the conditions of the chemical reaction. 
The detailed mechanisms of H$_{2}$O splitting and CO$_{2}$ reduction have been reported in several reviews, see, 
e.g., Ref.~\cite{FF}.

Figure~\ref{fig10}(b) shows the applicability of M-TNTs as catalysts for water splitting and carbon dioxide reduction. 
For water splitting, as already mentioned in the previous paragraph, the VBE has to be higher (more positive) than 
the  water oxidation potential O$_{ 2}$/H$_{ 2}$O (1.23 eV/NHE), and the CBE 
has to be lower (more negative) than the redox potential of H$^{ +}$/H$_{ 2}$ (0 eV/NHE). Therefore, 
the bandgap of the photocatalyst has to be larger than 1.23 eV ($\sim$ 1000 nm) to split water into H$_{ 2}$ and O$_{ 2}$, 
which is the minimum Gibbs free energy for this process. Here, the band edges are measured with respect to 
the normal hydrogen electrode (NHE), and their determination is discussed in detail in many publications, 
see, e.g., Refs.~\cite{FADL,fadl2} and references therein.
For the CO$_{2}$ reduction, the CBE has to be lower (more negative) than the redox potential of the 
natural fuel/CO$_{2}$. The positions of the band edges (which depend on the bandgap) are the main criterion 
for specifying a good photocatalyst for H$_{2}$O splitting or CO$_{2}$ reduction. 
 
Figure~\ref{fig10}(b) demonstrates that Fe-TNT is best for producing hydrogen (from water splitting), 
methane, and methanol (from CO$_{ 2}$ reduction). The Co- and Ru-TNTs have the lowest bandgaps, but they
are useful photocatalysts for generating methane fuel only. The Cr-TNT is the best candidate for 
producing all the considered clean fuels in this study. The other M-TNTs in Fig.~\ref{fig10}(b) show  
photocatalytic activities which enables them to split H$_{2}$O and reduce CO$_{2}$ reduction better
than pristine TNT.

\section{Summary}
\label{sec:summary}
In this work we systematically discussed the electronic, magnetic, and optical properties of titania 
nanotubes doped with 3$d$, 4$d$, and selected 5$d$ transition metals, in order to elucidate their potential
for various applications. Our study has been based on hybrid density functional theory, which is known for
leading to most accurate (in comparison to other DFT-based approximations) results. 
The stability of M-doped TNTs can be understood, to a large extent,
in terms of the ionic size and the electronegativity of the metal dopants.
The magnetic moments of doped TNTs depend on the number of outer shell $d$ electrons of the dopant
(up to about half-filling of the outer shell, i.e., for $d^{1}$ to $d^{6}$), and on the
coupling between the outer shell electrons (in particular, for higher fillings, $d^{7}$ to $d^{10}$).  

Dopant-related states at the Fermi energy for one spin component are found in several M-doped TNTs
(see Tables \ref{1}, \ref{2}, and \ref{3}), giving rise to ``spintronic'' properties. The metal dopants, except 
for Zr, create midgap states which implies a narrowing of the bandgap as compared to the pristine structure. 
Therefore all M-doped TNTs are more useful for optical applications than pristine TNT. The calculations
demonstrate that Mo-doped TNT has the highest optical activity as compared to other doped structures. 

The bandgap and the position of the dopant states determine the enhancement of the photocatalytic sensitivity.
In particular, Cr-, Mn-, Fe-, Co-, Nb-, Ru-, Ta-, and W-doped TNTs are expected to be preferential
for photocatalytic applications (water splitting and carbon dioxide reduction) as compared to pristine TNT. 
Fe-doped TNT is the best candidate for water splitting and for the production of hydrogen, methane, and 
methanol fuels, while the Cr- and W-doped TNTs are best for water splitting and CO$_{2}$ reduction, i.e.,
for the production of clean fuels and, at the same time, for helping to decrease the CO$_{2}$ pollution.
However, in several cases (14 out of 24) the created midgap states prevent an enhancement of 
photocatalytic sensitivity. Our results compare favorably with available experimental observations.

\acknowledgments{Financial support from the Deutsche Forschungsgemeinschaft (project number 107745057, TRR 80)
is gratefully acknowledged.}

\bibliography{bibfile}

\end{document}